# CRISTAL-ISE : PROVENANCE APPLIED IN INDUSTRY

Jetendr Shamdasani[1], Andrew Branson[1], Richard McClatchey[1], Coralie Blanc[1], Florent Martin[2], Pierre Bornand[2], Sandra Massonnat[3], Olivier Gattaz[3] and Patrick Emin[3]

[1]*CCCS Rearch Centre, Universityof the West of England, Coldharbour Lane, Bristol, United Kingdom*
[2]*Alpha-3i, Alpha-3i, 42 rue Rene Cassin, Rumilly, France*
[3]*M1i, 15 route de Nanfray, 74960 Cran Gevrier, France*
{jetendr.shamdasni,andrew.branson,richard.mclatchey,coralie.lucie.blanc}@cern.ch,{fmartin, pbornand}@alpha3i.com,{sandra.massonnat, olivier.gattaz, patrick.emin}@aglium.com

Keywords: Provenance, Semantic Web, Industrial Collaboration, XML, XML Schema, BPM, Resource Allocation

Abstract: This paper presents the CRISTAL-iSE project as a framework for the management of provenance information in industry. The project itself is a research collaboration between academia and industry. A key factor in the project is the use of a system known as CRISTAL which is a mature system based on proven description driven principles. A crucial element in the description driven approach is that the fact that objects (Items) are described at runtime enabling managed systems to be both dynamic and flexible. Another factor is the notion that all Items in CRISTAL are stored and versioned, therefore enabling a provenance collection system. In this paper a concrete application, called Agilium, is briefly described and a future application CIMAG-RA is presented which will harness the power of both CRISTAL and Agilium.

## 1 INTRODUCTION

The purpose of this position paper is to make the case for the use of provenance (Moreau, 2010) in industrial systems. Currently provenance gathering systems and techniques have mostly been used within scientific research domains such as neuroscience (Anjum et al, 2011) or other areas of bioinformatics (Goble, 2002). The usefulness of the application of this technology has been discussed at length elsewhere and is not the primary purpose of this paper. The interested reader is directed to other works such as (Simmhan et al, 2005). However, we have developed a concrete application of provenance management for industry which is discussed in this paper. The secondary purpose of this paper is to present the CRISTAL-iSE project.

CRISTAL-iSE (CRISTAL-iSE, 2013) is a project about fostering collaboration between industrial partners and academia and the development of individual researchers in the project. The partners involved in the project are from the University of the West of England (UWE, Bristol, United Kingdom, Academic), M1i (Annecy, France, Commercial) and Alpha-3i (Rumilly, France, Industrial). At the core of the project is a software product known as CRISTAL (Branson et al., 2013). This software had previously been developed at the European Organisation for Nuclear Research (CERN, Switzerland), in collaboration with UWE and the Centre National de la Recherche Scientifique (CNRS, France). At the end of the project it is foreseen that there will be three final pieces of software developed. These will be a new open source version of CRISTAL (UWE), a new version of Agilium (M1i) based on this version of CRISTAL and CIMAG-RA (Alpha-3i) which will be built upon both the work developed by UWE and M1i.

The main concept behind CRISTAL is what is known as a "description driven" system (Estrella, 2003). The main strength of such a "description driven" approach is that users who develop models of systems need only define them once to create a usable application. CRISTAL then orchestrates the execution of the processes defined in that model (with the consequent capture of provenance information). These descriptions can be modified at runtime and can capture almost any domain; this flexibility has been proven by its use in the construction of the CMS ECAL (CMS, 2008) at CERN. It has also been applied to the Business Process Management (BPM) domain (to model business-based process workflows) and the manufacturing domain (to control manufacturing processes) and is currently being applied to the HR domain allowing users to modify defined processes.

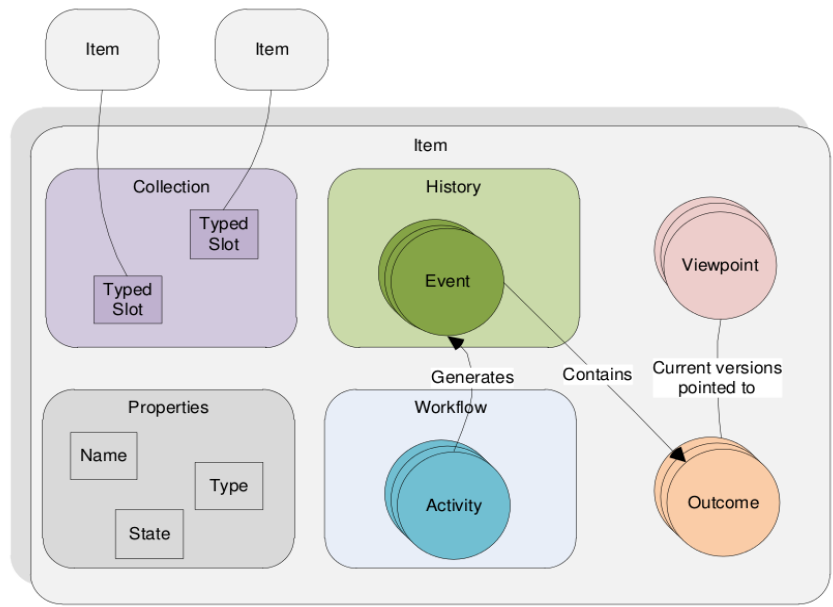

Figure 1: The internals of a CRISTAL Item.

Besides flexibility being a fundamental function of the system, another area where CRISTAL excels is the area of provenance. Provenance within the computer science literature is defined as the source or origin of an artefact or a piece of data. Within CRISTAL every defined element (or Item) is stored and versioned; therefore, nothing is ever deleted. This allows users of the system to view older versions of their Items (akin to Objects in Object Orientation) at a later date and either extend or return to a later version of an Item. A full description of the CRISTAL provenance model is out of the scope of this paper. Please see (Shamdasani, 2012) for further detail and (Branson et al., 2013) for a full description of the CRISTAL system. However, in this section the notion of an Item is briefly elaborated upon so that the content of this paper can be better understood.

There follows a very brief description of Item elements (see Figure 1) and their associations:

- Workflows are complete layouts of every action that can be performed on that Item, connected in a directed graph that enforces the execution order of the constituent activities.
- Activities capture the parameters of each atomic execution step, defining what data is to be supplied and by whom. The execution is performed by agents.
- Agents are either human users or mechanical/ computational agents (via an API), which then generate events.
- Events detail each change of state of an Activity. Completion events generate data, stored as outcomes. From the generation of an Event provenance information is stored.
- Outcomes are XML documents resulting from each execution (i.e. the data from completion Events), for which viewpoints arise.
- Viewpoints refer to particular versions of an Item's Outcome (e.g. the latest version or, in the case of descriptions, a particular version number).
- Properties are name/value pairs that name and type items. Properties also denormalize collected data for more efficient querying, and
- Collections enable items to be linked to each other.

The outcome of the CRISTAL-iSE project will be a newer, open source version of the CRISTAL Kernel, developed and managed by UWE, a new version of M1i's BPM use of CRISTAL known as Agilium and a Resource Allocation use of Agilium with CRISTAL known as CIMAG-RA developed and managed by Alpha-3i. As well as software outcomes the research focus of CRISTAL-iSE is to add Semantic and Distribution capabilities to the Kernel. The semantic work has already begun where the CRISTAL provenance model is being made compliant with the Open Provenance Model (OPM) (Moreau, 2010).

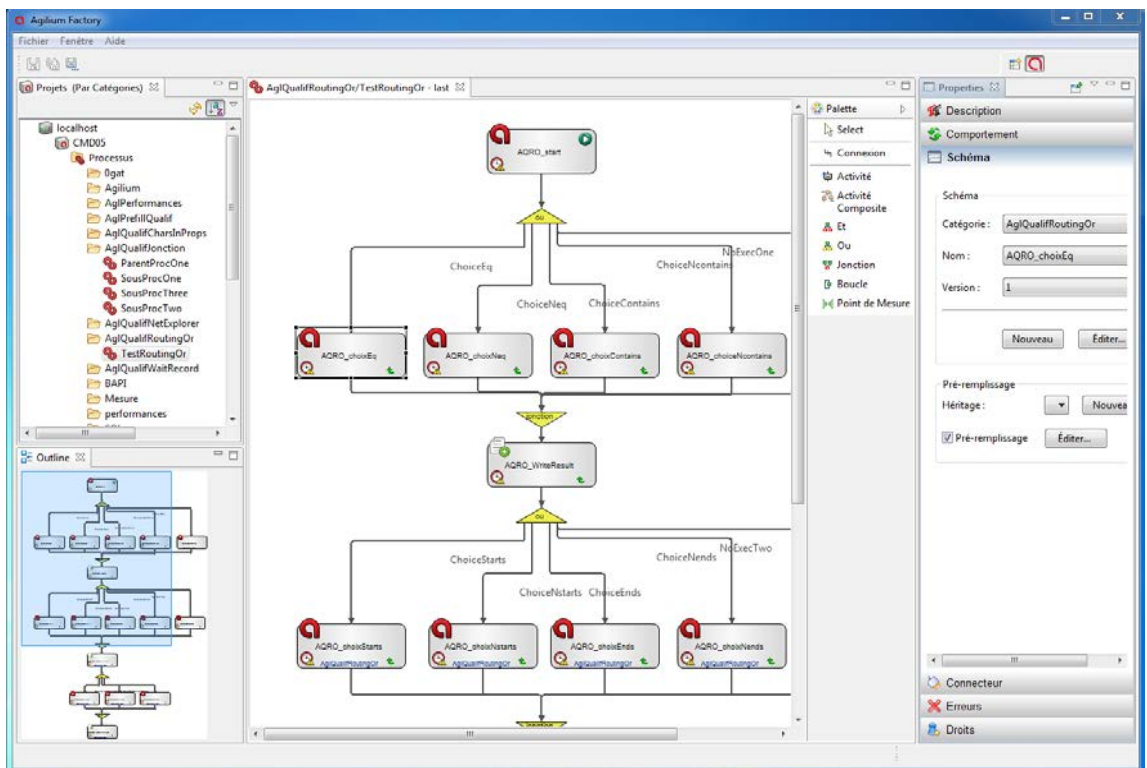

Figure 2: The Agilium Factory Application.

This paper is organised as follows, section 2 describes the current situation at M1i and its BPM specific use of CRISTAL. Section 3 describes the envisaged Alpha-3i application. Finally section 4 presents conclusions and future work by the CRISTAL-iSE partners.

## 2. CRISTAL IN AGILIUM

One successful commercial application of CRISTAL outside of CERN has been the Agilium product (Agilium, 2013). It is currently developed and sold to industrial and commercial clients via the M1i company. This is a BPM orientated system which consists of four main parts, one server application and three user interfaces.

The Agilium Server is based on CRISTAL, but with several domain extensions and support for additional protocols added. The user interface (UI) components are the Agilium Web component, the Agilium Supervisor GUI and the Agilium Factory. The Agilium Web is a web application based on J2EE and running within Tomcat as the container. This is where users can browse the currently active jobs and different instances of business processes. The list of jobs available to a user are constrained by their individual roles (for example, administrator). The web UI also allows users to complete manual activities.

The supervisor GUI component of Agilium is derived from the original Java Swing CRISTAL GUI, and is used by administrators of the system to be able to design and debug workflows and for general system management. The key component in Agilium is known as the *Factory*. The factory is a full Eclipse based application which has a modern UI and allows M1i's users to create and manage their own CRISTAL based workflows. A screenshot of the Agilium Factory is shown in figure 2.

The major area that Agilium uses CRISTAL for is provenance capture and recording of their BPM process executions. Within Agilium, the provenance model is identical to the provenance model of CRISTAL where Events are generated and stored. However, it is applied to the domain of BPM. As stated previously, all models are created at runtime. This means that all BPM workflows developed within Agilium are stored and versioned. This allows users to come back at a later date and view previous versions of the BPM models, fix bugs, or extend their previous BPM workflows.

As example of where provenance is useful for Agilium is a company which produces solar panels. With this client, the production of each solar panel can take more than a month. They also require different versions of workflows to be stored and accessed on site. Therefore, this client of theirs requires that they be able to look into the *past* versions of their processes and workflows. This means that they can retrieve the history of all the production steps for each panel, even though the BPM workflow has evolved between the two generations of panels.

As an example of an alteration to the fabrication process, in the past they have modified their production process to increase the performance level of the solar cells. The workflows corresponding to the production processes are modified to add or remove activities matching an electro-deposition or cleaning step, or alter their parameters. These modifications are usually done at run time. Therefore, these changes are saved and stored as newer versions, thereby, allowing the panels using the older versions of the workflow to continue unhindered whereas the newer modifications can be applied to newer solar panels in production; this is a key strength of using CRISTAL in Agilium and demonstrates not just the use of provenance but also the flexibility of the system.

The inherent provenance capabilities of CRISTAL mean that the model itself is also versioned, allowing users to look at the production steps for each version of the panels they have created and to see what processes they have in common. This also allows them to view and analyse which processes have changed. This aspect is crucial to their business since it allows them to look at the evolution of the production process. The developers at M1i chose CRISTAL as the basis for their system since they felt that its provenance and traceability features were key for them to create a product with a competitive edge in the market.

## 3. ALPHA-3I AND CIMAG-RA

CIMAG-RA is a resource allocation and management solution (May, 2012) which the Alpha-3i company will market to clients as an outcome of the CRISTAL-ISE project. This solution uses both the Agilium component (described previous) and CRISTAL. They require a solution which can track and predict resource allocation to tasks over time. There are different types of resources that have been identified by Alpha-3i during the course of their investigation, the different categories are: Machines (Physical, mechanical non human entities), Operators (Human entities that operate machines) and Administrators (Human entities that control the system). These resources are involved in the completion of certain tasks. Each resource that has to perform a task must have a correct set of properties. Therefore there must be a matching of properties to tasks to be completed. This is the resource allocation problem that requires solution.

A key factor of the CIMAG-RA solution is the requirement for provenance to influence future scheduling. This entails the storage of CRISTAL Events that occur on their defined Item types (human, machines or tasks). The use of this information will aid in optimizing customer processes and provide useful information to administrators. In detail, provenance is used for different purposes:

- To gauge the relevance of an allocation of a resource to a specific task. This means storing the experiences of a human along with the set of tasks they have processed in the past, can help with the decision making process. Therefore here provenance information can be used as data which helps users gauge the trustworthiness of the current decision making process.
- To create an audit trail or provide information to resolve disputes. For example, in healthcare applications, this would be data containing the healthcare history of a single patient, the procedures carried out on that patient and the resources (nurses, doctor's machines etc.) allocated to each procedure performed on a patient. This information must be stored in retrievable fashion for legal or analysis purposes (Wang et al, 2007).
- To optimize non-human resource management. For example, in the manufacturing industry, storing and then analysing the evolution of a machine resource can be a key factor to determine the lifecycle of an instrument and may aid in predicting when it will need repair or replacement.

Another requirement of the CIMAG-RA solution is flexibility. Indeed to meet requirements of a large variety of clients from heterogeneous markets such as manufacturing, medical, financial services and others, the solution must be fully reconfigurable without recompiling code. More precisely, resources and tasks will be designed and modified according to client needs or legal constraints. Therefore, a crucial requirement is to track objects and their provenance over time.

Many different resource management solutions exist today (Vanden Berghe, 2002 and Lombardi, 2012), however, after some investigation the developers at Alpha-3i found that these did not fit their needs. Mostly due to the requirements of provenance and flexibility, described earlier as being novelties in CRISTAL, are crucial requirements for the CIMAG-RA software.

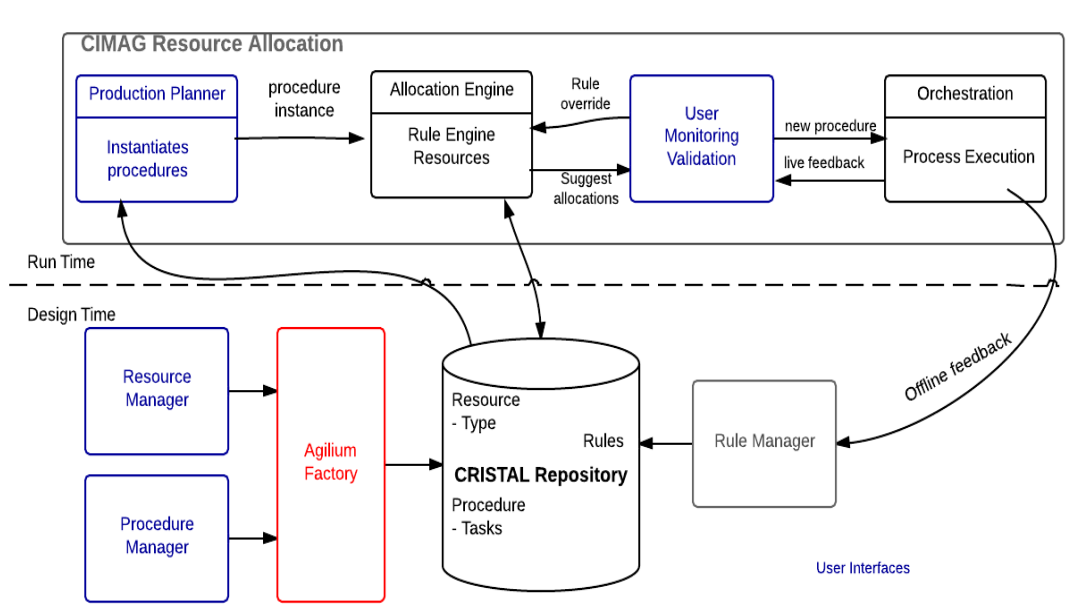

Figure 3: The CIMAG-RA Application Architecture.

Figure 3, shows the architecture of CIMAG-RA and how they use the CRISTAL kernel. It is foreseen that Alpha-3i will implement, using CRISTAL, a repository of procedures, i.e. a set of tasks, and resources, where changes are tracked to the resource and procedure management schemes and then stored within the database of choice. From a CRISTAL point of view, procedures, tasks and resources are Items described by their own schema. Such Items can evolve, thus the repository has to manage Items that are an instance of an object (resource, procedure and tasks) corresponding to a schema Item regarding its version.

At design time, the Agilium factory is used to define the resources and procedures the same way Agilium users define their BPM Items. Concretely, we use the Resource and Procedure Manager components (left of figure 3) to design the schema Items of our applications. Such schema Items are then instantiated as resource and procedure Items and populated using the data and constraints of the client. For example, in the manufacturing area, a named operator resource cannot work more than 12 hours per day and he will be on holiday from Monday to Wednesday next week. So these and other constraints will populate his corresponding Item and therefore he will be matched to a corresponding task based on other requirements.

At run time, the "Allocation Engine" component retrieves the resources and procedures as defined at design time, to provide an output to the planner. The "Allocation Engine" is a specific activity that takes inputs (resources and tasks) and processes them using a script. Its output is the allocation of resources to tasks and more globally procedures. As a result, this impacts the status of resources and procedures and consequently the content of corresponding Items in the repository. So resources and task Items are updated in the repository using CRISTAL. The retrieval and update services are represented by the double arrow between repository and allocation engine.

Finally the output is validated by the planner and run by the "Orchestration" module. The Orchestration module is in charge of both executing and monitoring the procedures regarding the allocation model and providing feedback to planner and the rule manager module. Regarding the feedback from Orchestration module, the planner can manually update the output from allocation engine. This can also be done automatically by the Rule manager module itself. These operations have also an impact on the Items in repository. This simple example illustrates the use of provenance capture and management in the CRISTAL empowered version of CIMAG-RA.

## 4. CONCLUSIONS AND FUTURE WORK

This paper has introduced the CRISTAL-iSE project and outlined its aims and objectives. The major focus of the project is the collaboration between academia (UWE) and industry (M1i and Alpha-3i). During this collaboration the researchers in the project have been able to demonstrate uses for CRISTAL and its flexibility, particularly in the area of provenance exploitation in commercial applications. Thus the main research focus of the project is the use of provenance within Industry.

One application of CRISTAL has already been presented, where CRISTAL has been converted to a system for BPM use M1i's Agilium). They have been collecting data from clients for over ten years. A potential and *new* application of CRISTAL is currently emerging with Alpha-3i's CIMAG-RA application. In this application, both the new versions of Aglium and CRISTAL will be used to create a Resource Allocation application in the Human Resources (HR) domain.

From the large datasets that are available already, an OPM (Open Provenance Model, Moreau 2010) compliant provenance model will be created to foster collaboration between the wider provenance research communities. However, this work is currently on going and will be demonstrated at a later date.

Currently within the project an initial requirements gathering exercise has been completed and initial designs have been created to move forward with the applications that should arise from the end of the project. These requirements have led to a more "modular" design of the CRISTAL system with allowing a generic core or kernel to be available to the wider community.

From a functional point of view, the CRISTAL kernel as is, allows the management of Items such as process activities or tasks, workflows or procedures, resources and scripts as defined in previous sections. It also provides provenance capabilities and flexibility. To fulfil the Alpha-3i requirements, we need first to define resources based on interoperable standards such as HR-XML and ISA 95. This approach will ease integration with third-party applications. We will then implement a rule based engine to provide a logic module on top of the kernel.

## ACKNOWLEDGEMENTS

This project has been funded the Marie-Curie Industrial and Academic Partnership Scheme (IAPP) scheme. The authors would like to thank their home organisations and, in particular, Becky Gooby and Bruno Malagola for their efforts in contributing to the project.